\documentclass[
 reprint,
 amsmath,amssymb,
 aps,
 showkeys,
prb,
floatfix,
]{revtex4-2}
\usepackage{graphicx}
\usepackage{dcolumn}
\usepackage{bm}
\usepackage[colorlinks=true,linkcolor=blue,citecolor=blue,urlcolor=blue]{hyperref}
\usepackage[T1]{fontenc}
\usepackage[utf8]{inputenc}
\usepackage{subdepth}
\usepackage{physics}
\usepackage{xcolor}
\usepackage{natbib}

\usepackage{mathtools}
\mathtoolsset{showonlyrefs}

\usepackage[normalem]{ulem}

\usepackage{fancyhdr}
\DeclareMathOperator{\erfc}{erfc}
\allowdisplaybreaks

\begin{document}

\title{Exciton Diffusion in a Quantum Dot Ensemble}

\author{Karol Kawa}
\email{karol.kawa@pwr.edu.pl}
\author{Pawe{\l} Machnikowski}
\affiliation{Institute of Theoretical Physics, Wroc{\l}aw University of Science and Technology, 50-370 Wroc{\l}aw}

\date{\today}

\begin{abstract}
In this theoretical study, we explore F\"orster resonant energy transfer of a single exciton within a two-dimensional array of self-assembled quantum dots arranged randomly on a circular mesa.
Employing the stochastic simulation method, we solve the equation of motion for the density matrix, considering a specified decay rate.
Our analysis quantifies diffusion through the mean-square displacement from the initially excited quantum dot, revealing distinct temporal stages: ballistic, normal diffusion, and saturation.
Furthermore, we observe power-law localization of the exciton.
Complementing our numerical investigations, we develop approximate analytical expressions that closely align with the numerical findings.
\end{abstract}

\maketitle
\thispagestyle{plain}
\section{\label{sec:introduction}Introduction}

F{\"o}rster resonance energy transfer (FRET) appears widely in the quantum world, both in biological and technical structures.
The former include light harvesting systems that employ cascade-like transport to move solar energy from pigment molecules to the reaction centers of photosynthesis \cite{Mirkovic2016,Ritz2002}.
The latter are dominated by exciton diffusion in different types of quantum dots (QDs), e.g., QD solids \cite{Kholmicheva2016}, nanocrystals \cite{santos_quantum_2020,kulak_nonradiative_2022}, self-assembled QDs \cite{DeSales2004, Kozub2012EnhancedCoupling++, Miftasani2016}, and silicon QDs \cite{Cao2019}.
In particular, in Ref.~\cite{DeSales2004} excitation transport has been observed in self-assembled QD ensembles with varying planar density.
In that work, transport even beyond the excitation region was confirmed by the observed spatial width of photoluminescence (PL) that exceeded the excitation laser spot.
Furthermore, an enhanced effect was observed in the case of diluted samples, eliminating the hypothetical possibility of carrier tunneling (Dexter transfer).

Energy transport in a QD ensemble involves the diffusion of an exciton, i.e. an~electron-hole pair initially formed (e.g.~optically) in individual QDs.
The resonant energy transfer mechanism proposed by F{\"o}rster \cite{F_rster_1948,F_rster_1959,Lehmberg1970RadiationSystem} offers an explanation of transport in such systems.
The excited donor-QD transfers its excitation to the remaining acceptor-QDs coupled via dipole interactions.
Assuredly, the transport mechanism (F\"orster vs. Dexter) can be experimentally determined by analyzing optical spectra in coherent two-dimensional spectroscopy \cite{Specht2015}.

The exact formula for Forster's couplings in a planar set of dipole emitters (e.g.~self-assembled QDs) is known.
It was derived by transforming the minimal coupling Hamiltonian in a dipole approximation using the Power-Zienau-Wooley (PZW) transformation \cite{Power1959-fd,Woolley1971-dw,Cohen-Tannoudji2008-ms}.
The resulting dipole-dipole power-law coupling has the form of~a~sum of~three long-range terms decreasing with the~distance, each multiplied by an oscillating factor \cite{Lehmberg1970RadiationSystem, Stephen1964,Kozub2012EnhancedCoupling++,Miftasani2016}.

Self-assembled QDs are not perfectly homogeneous (due to~randomness of~the~growth process).
Therefore, QDs differ in the fundamental transition energy of forming an electron-hole pair.
Such an energy dispersion makes the set of QDs a disordered system similar to that described by the Anderson tight-binding model \cite{Anderson1958,Rodriguez2003AndersonHopping} with long-range hopping integral.
Although most of the research on Anderson's localization in disordered systems focused on nearest-neighbor-coupled models, Anderson's original work dealt with long-range couplings, decreasing with some power of the inter-site distance, $V(r)\propto r^{-\mu}$.

In this paper, we theoretically investigate the diffusion of a single exciton in a planar QD ensemble restricted to a circular mesa.
Exciton diffusion is a manifestation of FRET that stems from long-range dipole-dipole couplings.
The fundamental transition energy disorder has a negative impact on the range and speed of diffusion.
A general framework for describing (quasi)particle diffusion in such systems was proposed in~Ref.~\cite{Kawa2020DiffusionCoupling}, where we investigated the diffusion of a single excitation in a regular one- or two-dimensional lattice with strong on-site disorder and inter-site coupling that decreases inversely proportional with distance $V(r)\propto r^{-1}$. 
Here, we extend the considerations to a realistic system: a planar ensemble of randomly placed self-assembled quantum dots with fundamental transition energy disorder, coupled by the long-range oscillating F{\"o}rster couplings.
We also assume a finite exciton lifetime and compare it to the idealized non-dissipative case.
To simulate the dissipation process, we employ the numerical method of stochastic simulations, also known as the quantum jump method \cite{Breuer2002, Gardiner2010}.

The diffusion of an exciton is characterized by the mean-square displacement of the exciton as a function of time.
We show that this quantity evolves in three steps: first ballistic motion, then standard diffusion, and finally saturation (cf.~Ref.~\cite{Kawa2020DiffusionCoupling}).
Similarly, the growth of the exciton density at a given QD follows three steps: quadratic in time, followed by linear in time ending in saturation.
At the same time, the dependence of exciton density on the distance reveals power-law localization of the exciton in the system.
We explain the three-step diffusion and power-law localization using a model in which the disorder expressed by the standard deviation of the transition energies is much greater than the coupling strength between sites \cite{Anderson1958, Kawa2020DiffusionCoupling, Kawa2021SpreadOfCorrel}.
In such a model system, only first-order (direct) jumps from the excited site even to the remote ones are relevant.
This regime is opposite to the nearest-neighbor coupled systems.
By neglecting all the couplings except for those involving the initially occupied QD we are able to propose an approximate analytical solution to the exciton dynamics which, at least qualitatively, reproduces the simulation results.
This allows us to gain a deeper understanding of the transport mechanism.

The organization of this paper is as follows.
Sec.~\ref{sec:system_and_model} contains a detailed description of the investigated system, the model that describes it, and the quantum jump method used for numerical simulations of the system dynamics.
In Sec.~\ref{sec:numerical_results}, we present the exciton dynamics obtained from the numerical evaluation of the introduced model.
In Sec.~\ref{sec:analytic}, we propose an approximate analytical solution that reproduces the exciton dynamics.
Finally, in Sec.~\ref{sec:discussion} we conclude and discuss the results.

\section{System, Model and Simulation Method\label{sec:system_and_model}}
\begin{figure}
    \centering
    \includegraphics[width=0.8\linewidth]{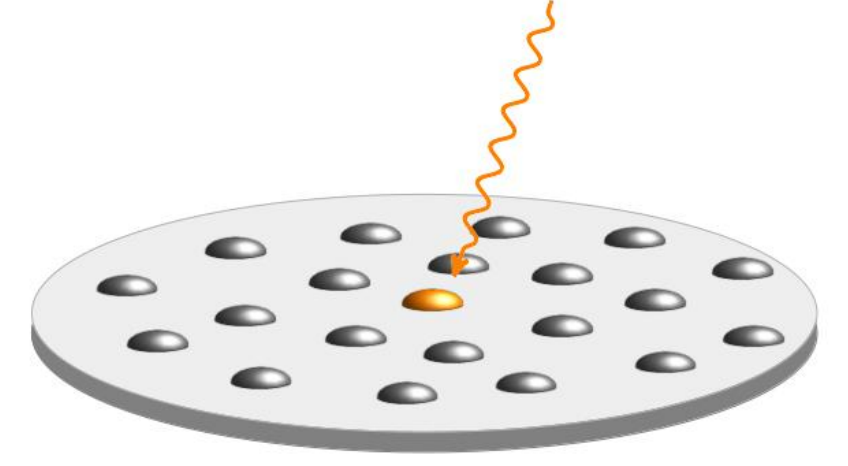}
    \caption{Schematic representation of the investigated system: planar ensemble of self-assembled QDs on a circular mesa.
    The growth axis is perpendicular to the plane of the ensemble.
    QD lying closest to the center is initially excited.}
    \label{fig:system}
\end{figure}

The system under study is a planar ensemble of self-assembled QDs randomly placed in the plane perpendicular to the growth axis (see Fig.~\ref{fig:system}) with constant surface density $\rho_A = 10^{11}$~QDs$/$cm$^2$.
In our model, QDs occupy a circular area of radius $R$.
Due to the constant planar density, the number of QDs is adjusted to the linear size of the mesa and is approximately equal to $N=\rho_A \pi R^2$.
That is, in the simulations, the number of QDs and the planar density $\rho_A$ are given, and the size of the mesa is adjusted.
The average distance between the~nearest sites is $r_\mathrm{av} \approx (\rho_A)^{-1/2} \approx 32$~nm.
However, the~minimum distance between QDs is limited to~$d_\mathrm{min} = 10$~nm, which roughly corresponds to~the~minimum diameter of~a~single QD.
The positions of~the~dots are denoted by~$\bm{r}_\alpha$.
The investigated system is not homogeneous.
Quantum dots differ in their fundamental transition energy.
We denote this energy by $E_\alpha = \overline{E} + \epsilon_{\alpha}$ for the $\alpha$-th QD, where $\overline{E}$ corresponds to its average, which for exemplary CdTe QDs is approximately $2.59$~eV, and $\epsilon_\alpha$ is a small deviation from the average, modeled here by a symmetric normal distribution $\mathcal{N}(0,\sigma^2)$ of variation $\sigma^2$.
The value of $\sigma$ for CdTe grown with the Stranski-Krastanov process is typically $\sim 50$~meV, however, here we consider much more homogeneous systems.
Highly uniform QDs can be fabricated by local droplet etching~\cite{Heyn2009}, which nowadays achieves narrow emission linewidths of the order of $\mu$eV \cite{daSilva2021,Zhai2020} for a single QD and less than 10 meV for ensembles.
Our considerations apply to both the Stranski-Krastanov and droplet-etched self-assembled systems.

Although the size of a single QD is much smaller than the relevant radiation wavelength, the size of the ensemble can exceed it several times.
Thus, the dipole approximation cannot be applied to the system as a whole.
Instead, we employ the PZW transformation \cite{Power1959-fd,Woolley1971-dw,Cohen-Tannoudji2008-ms}.
It converts the minimal-coupling Hamiltonian in~the~Coulomb gauge representing an ensemble of small physical systems into the Hamiltonian of quantum emitters treated as a point dipole each, coupled to the electric displacement field $\bm{\hat{D}}(\bm{r})$.

Then, the QD ensemble, together with the surrounding radiation and the mutual interaction is described by the Hamiltonian,
\begin{gather}
    \hat{H} = \hat{H}_\mathrm{at} + \hat{H}_\mathrm{rad} + \hat{V}_{\textrm{at-rad}},
    \label{eq:hamiltonian-after_PZW-transfromation}
\end{gather}
where
\begin{equation}
    \hat{H}_\mathrm{at} = \sum_\alpha \epsilon_\alpha \hat{\sigma}^\dagger_\alpha \hat{\sigma}_\alpha
\end{equation}
is the Hamiltonian of the dipole emitters (QDs, i.e., artificial atoms) with $\hat{\sigma}_\alpha$ representing the transition operator that annihilates the excitation at the site $\alpha$.
The second term corresponds to the photon bath,
\begin{gather}
    \hat{H}_\mathrm{rad} = \sum_{\bm{k}\lambda} \hbar \omega_k \hat{b}^\dagger_{\bm{k}\lambda} \hat{b}_{\bm{k}\lambda},
\end{gather}
where $\hat{b}_{\bm{k}\lambda}$ ($\hat{b}_{\bm{k}\lambda}^\dagger$) is the annihilation (creation) operator for photon of wave vector $\bm{k}$ and polarization $\lambda$, whereas $\omega_k$ is the corresponding photon frequency.
The last term,
\begin{gather}
    \hat{V}_\textrm{at-rad} = -\frac{1}{\varepsilon_0 \varepsilon_r}\sum_\alpha \bm{\hat{d}}_\alpha \cdot \bm{\hat{D}} (\bm{r}_\alpha),
    \label{eq:coupling_PZW_transform}
\end{gather}
stems from the PZW transformation and corresponds to the coupling between the QDs and the electromagnetic field.
Here, $\bm{\hat{d}}_\alpha = \bm{d}_0 \hat{\sigma}_\alpha + \bm{d}^*_0 \hat{\sigma}_\alpha^\dagger$ is the dipole moment operator for the emitter $\alpha$, while $\varepsilon_0$ and $\varepsilon_r$ are the electric permittivity of the vacuum and material, respectively.
The dipole operator in question corresponds to the creation of an exciton with a given angular momentum of $+1$ or $-1$ via circularly polarized laser beam [Fig.~\ref{fig:system}].
Here, we limit ourselves to assuming constant exciton polarization throughout the entire evolution.
The justification for such a simplification is presented in Appendix~\ref{appendixB}, where we discuss exciton diffusion in an enhanced model, addressing both bright exciton states and exciton fine structure splitting (FSS).
Neither spin-flipping F\"orster transfer nor exciton FSS significantly impact the exciton transport.

The displacement field $\bm{\hat{D}}(\bm{r})$ is expressed by the photon operators as
\begin{gather}
    \bm{\hat{D}}(\bm{r}) = i~\sum_{\bm{k}\lambda} \sqrt{\dfrac{\hbar \varepsilon_0\varepsilon_r \omega_k}{2\mathcal{V}}} \hat{e}_{\bm{k}\lambda}\hat{b}_{\bm{k}\lambda} e^{i\bm{k}\cdot \bm{r}} + \mathrm{H.c.},
\end{gather}
where the unit vector $\hat{e}_{\bm{k}\lambda}$ determines the polarization of light in the mode with wave vector $\bm{k}$ and polarization $\lambda$, and $\mathcal{V}$ is the normalization volume.

The equation of~motion for~the~reduced density matrix $\rho$ can be derived following the steps of~Ref.~\cite{Lehmberg1970RadiationSystem}.
First, one looks at~the~evolution of~the quantum mechanical average of any operator $Q$ in the Heisenberg picture.
The evolution is governed by a Hamiltonian, which includes both electronic and~photonic degrees of~freedom.
The~equations of~motion for~the atomic and photonic operators are found and the latter are then eliminated, leading to an integro-differential equation for the atomic operators which is reduced by using the Markov approximation.
In~the~end, one neglects off-resonant terms and radiation-induced energy shifts and~rewrites the~equation in the Schr\"odinger picture obtaining the equation of motion (see also Refs.~\cite{Miftasani2016,Kozub2012EnhancedCoupling++}),
\begin{equation}
    \frac{\dd}{\dd t}\hat{\rho} = -\dfrac{i}{\hbar}\left[\hat{H}_0,\hat{\rho}\right] + \sum_{\alpha\beta} \Gamma_{\alpha\beta} \left(\hat{\sigma}_\alpha \hat{\rho} \hat{\sigma}_\beta^\dagger - \dfrac{1}{2}\left\{\hat{\sigma}^\dagger_\beta \hat{\sigma}_\alpha,\hat{\rho}\right\}\right),
    \label{eq:master_equation}
\end{equation}
where
\begin{equation}
    \hat{H}_0 = \hat{H}_\mathrm{at} + \sum_{\alpha\beta} V_{\alpha\beta}\hat{\sigma}^\dagger_\alpha \hat{\sigma}_\beta.
    \label{eq:hamiltonian_atomic_plus_coupling}
\end{equation}
The first term on the right-hand side of Eq.~\eqref{eq:master_equation} corresponds to the unitary dissipationless evolution of the system, whereas the second term is the Linblad part responsible for the dissipation process.
The long-range coupling between the QDs is expressed by
\begin{equation}
    V_{\alpha\beta} = \hbar \Gamma_0 G(k_0 r_{\alpha\beta}),
    \label{eq:coupling}
\end{equation}
with $V_{\alpha\alpha}=0$, where $\bm{r}_{\alpha\beta} = \bm{r}_\alpha - \bm{r}_\beta$, $\Gamma_0=|d_0|^2 k_0^3/(3\pi \varepsilon_0 \varepsilon_r)$ is the spontaneous emission rate for a single dot and $k_0=2\pi n_\mathrm{refr}/\lambda_0 =  n_\mathrm{refr}\overline{E}/(\hbar c)$, where $n_\mathrm{refr}$ is the refractive index of the medium and $c$ denotes the speed of light.
The values of these and other material parameters are gathered in Table~\ref{tab:parameters}.
The coupling has a mixed power-law and oscillating character~\cite{Lehmberg1970RadiationSystem},
\begin{equation}
    G(x) = -\dfrac{3}{8}\left(\dfrac{\cos x}{x}+\dfrac{\sin x}{x^2} + \dfrac{\cos x}{x^3}\right).
    \label{eq:coupling_function}
\end{equation}
The short-range couplings provided by the overlap of the QDs wave functions and/or Coulomb correlations were neglected due to relatively large interdot distance (wave functions do not overlap).

The second term in Eq.~\eqref{eq:master_equation} is responsible for the dissipation process.
The dissipator coefficients are $\Gamma_{\alpha\alpha} = \Gamma_0$ and
\begin{equation}
\Gamma_{\alpha\beta}=\Gamma_0 F(k_0 r_{\alpha\beta}) \text{ for }\alpha\neq\beta, 
\label{eq:dissipator}
\end{equation}
 where
\begin{equation}
    F(x) = \dfrac{3}{4}\left(\dfrac{\sin x}{x} - \dfrac{\cos x}{x^2} + \dfrac{\sin x}{x^3}\right),
    \label{eq:dissipator-distance-dependence}
\end{equation}
cf. Refs.~\cite{Kozub2012EnhancedCoupling++,Miftasani2016}.

\renewcommand{\arraystretch}{1.35}
\begin{table}[tb]
\begin{tabular}{l c c l}
\hline\hline
Description & Symbol & Value & Unit \\
\hline
Spontaneous emission rate           & $\Gamma_0$  & 1.0/0.39 & ns$^{-1}$\\
Resonance wave length (vacuum)      & $\lambda_0$ & 479 & nm\\
Refractive index                    & $n_\mathrm{refr}$ & 2.6 & - \\
Surface density of QDs (2D)         & $\rho_A$  & $10^{11}$ & cm$^{-2}$\\
Minimal QD diameter                 & $d_\mathrm{min}$ & 10 &  nm\\
\hline\hline
\end{tabular}
\caption{Material parameters of the system.}
\label{tab:parameters}
\end{table}

To efficiently simulate systems of thousands of dots, we employ the quantum jump method \cite{Breuer2002,Gardiner2010}.
In this approach, we consider a state vector of $N$ elements instead of a density matrix of $N^2$ elements, which unburdens the computational load.
The time-dependent state of the system $\ket{\Psi(t)}$ is expressed in the basis of the exciton located at a single dot $\{\ket{\alpha}\}_{\alpha=0}^{N-1}$,
\begin{gather}
\ket{\Psi(t)} = \sum_\alpha c_\alpha(t) \ket{\alpha},
\label{eq:ketPsi}
\end{gather}
where $c_\alpha(t)$ is the time-dependent probability amplitude for finding the exciton at the site $\alpha$.
For dissipationless systems, one can reduce the equation of motion [Eq.~\eqref{eq:master_equation}] to the first term, which yields a Schr{\"o}dinger equation that can be solved by exact diagonalization of the Hamiltonian~\eqref{eq:hamiltonian_atomic_plus_coupling}, see Ref.~\cite{Kawa2020DiffusionCoupling}.

\subsection*{Stochastic simulation method}
\begin{figure}[tb]
    \centering
    \includegraphics[width=\linewidth]{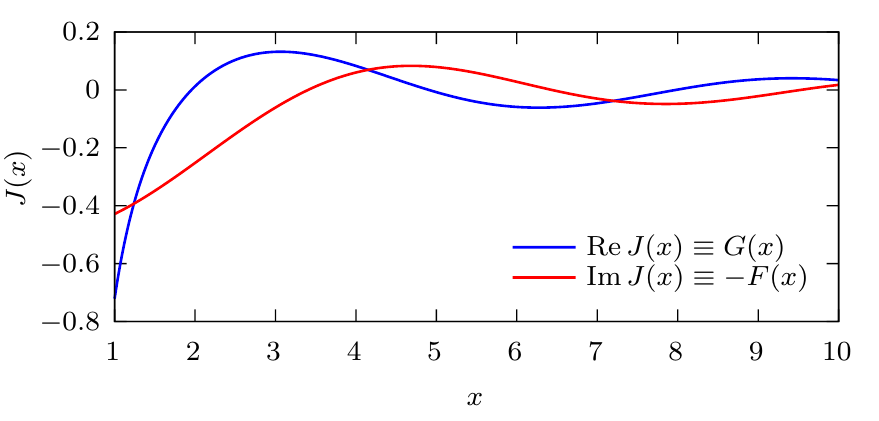}
    \caption{Plot of real and imaginary part of function $J(x)$ [Eq.~\eqref{eq:H(x)}].
    Its real part corresponds to function $G(x)$ [Eq.~\eqref{eq:coupling_function}] and its imaginary part corresponds to negative $F(x)$ [Eq.~\eqref{eq:dissipator-distance-dependence}]}.
    \label{fig:effective-potential-H-function}
\end{figure}

Now we briefly summarize the stochastic simulation method.
A detailed description of this method can be found in~Refs.~\cite{Breuer2002,Gardiner2010}.
The method is based on the conversion of the master equation [Eq.~\eqref{eq:master_equation}] into a piecewise continuous stochastic process.
Let us start by transforming Eq.~\eqref{eq:master_equation} into the equivalent form,
\begin{gather}
    \frac{\dd}{\dd t}\hat{\rho} = -\frac{i}{\hbar} \left(\hat{H}_\mathrm{eff}\hat{\rho} - \hat{\rho} \hat{H}_\mathrm{eff}^\dagger\right) + \sum_{i=1}^N \tilde{\Gamma}_i \tilde{\sigma}_i \hat{\rho} \tilde{\sigma}_i^\dagger,
\end{gather}
where
\begin{gather}
    \hat{H}_\mathrm{eff} = \hat{H}_0 + \frac{\hbar}{2i} \sum_{\alpha\beta} \Gamma_{\alpha\beta} \hat{\sigma}_\alpha^\dagger \hat{\sigma}_\beta
\end{gather}
is the effective non-Hermitian Hamiltonian which governs the stochastic evolution between jumps and where $\tilde{\Gamma}_i$ are the eigenvalues of the matrix $\Gamma_{\alpha\beta}$.
The corresponding eigenvectors are denoted by $u_i = (u_{i,\alpha})_{\alpha=0}^{N-1}$ and 
\begin{gather}
    \tilde{\sigma}_i = \sum_\alpha u_{i,\alpha}^* \hat{\sigma}_\alpha. 
\end{gather}
Here we also introduce the effective complex F\"orster coupling $V_{\alpha\beta}^\mathrm{eff} = V_{\alpha\beta} + \frac{\hbar}{2i}\Gamma_{\alpha\beta}$, for which $\hat{H}_\mathrm{eff} = \hat{H}_\mathrm{at} + \sum_{\alpha\beta} V_{\alpha\beta}^\mathrm{eff} \hat{\sigma}_\alpha^\dagger \hat{\sigma}_\beta$.
For $\alpha\neq\beta$ it is expressed as
\begin{equation}
    V_{\alpha\beta}^\mathrm{eff} = \hbar J(k_0 r_{\alpha\beta}),
    \label{eq:effective-potential-H-function}
\end{equation}
where
\begin{equation}
    J(x) = -\frac{3}{8} e^{ix} \left( \frac{1}{x} - \frac{i}{x^2} + \frac{1}{x^3} \right).
    \label{eq:H(x)}
\end{equation}
Real and imaginary part of function $J(x)$ is depicted in Fig.~\ref{fig:effective-potential-H-function}.
It will be useful in the following.

The unnormalized state $\ket{\Psi(t)}$ [Eq.~\eqref{eq:ketPsi}] evolves according to the Schr{\"o}dinger equation 
\begin{gather}
    i\hbar\frac{\dd }{\dd t} \ket{\Psi(t)} = \hat{H}_\mathrm{eff} \ket{\Psi(t)}
    \label{eq:continuous-evolution-between-jumps}
\end{gather}
until the jump occurs.
The time interval between jumps is a random variable with a cumulative distribution function $\mathcal{F}(t) = 1 - \braket{\Psi(t)}$.
Although continuous evolution preserves the number of excitons, each jump corresponds to the emission of a single photon.
In the case of multiple excitations present in the system, many jumps divided by continuous processes can occur.
Here, we restrict ourselves only to the single exciton initial state, which prohibits further evolution after the first jump.
The evolution of state $\ket{\Psi(t)}$ is calculated numerically according to Eq.~\eqref{eq:continuous-evolution-between-jumps}.
At each time step, one checks if the jump is to occur by comparing the current time with the jump time drawn from the distribution $\mathcal{F}(t)$.
At a jump, the state vector is transformed according to
\begin{gather}
    \ket{\Psi} \longrightarrow \frac{\sqrt{\tilde{\Gamma}_i}\tilde{\sigma}_i \ket{\Psi}}{\bigl|\sqrt{\tilde{\Gamma}_i} \tilde{\sigma}_i \ket{\Psi} \bigr|^{1/2}},
\end{gather}
which in our case (single exciton) is just $\ket{\Psi} \rightarrow 0$.
The simulation is performed multiple times for different disorder realizations of the energy and positions of QDs, as well as of the random jump process.
The desired quantity is then averaged over the repetitions.
\begin{figure}[t]
    \centering
    \includegraphics[width=\linewidth]{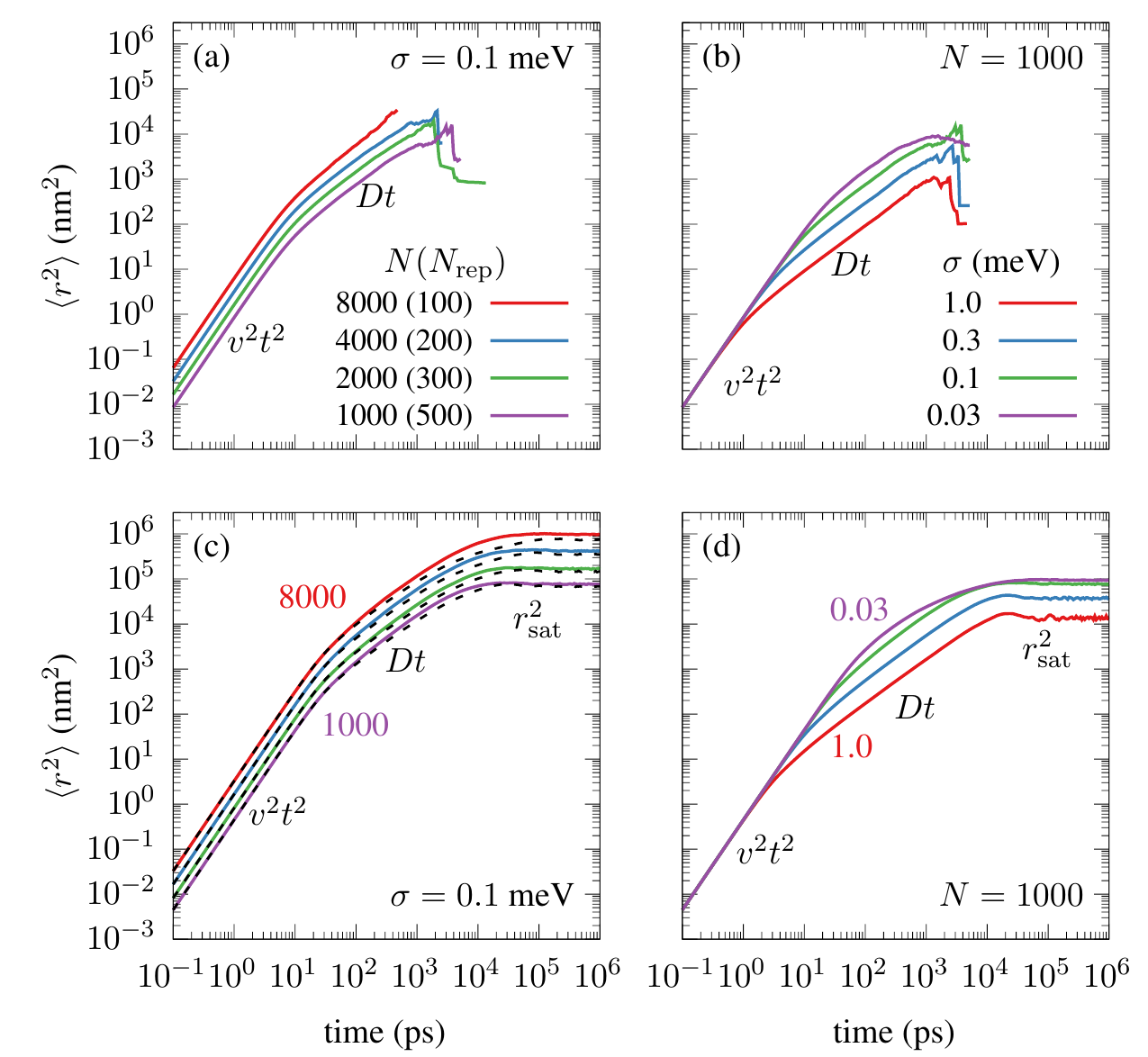}
    \caption{Mean square displacement of the exciton as a function of time for dissipative (a,b) and idealized nondissipative system (c,d) for different values of the system size and disorder strength. The number of realizations $N_\mathrm{rep}$ is given in the legend of panel (a) in parentheses for each system size.}
    \label{fig:evolution}
\end{figure}

\subsection*{Quantities describing the dynamics}
For a quantitative characterization of the diffusion, we are first interested in the spatial and temporal dependence of the occupation density, $\varrho_{\Delta r}(t,r)$, which is a normalized histogram of the occupation $\langle |c_\alpha(t)|^2\rangle$ where in each interval $\Delta r$ we count the occupations of QDs lying between $r$ and $r+\Delta r$, with fixed $\Delta r=5$~nm.
$\langle\dots\rangle$ corresponds to the average of $N_\mathrm{rep}$ realizations in which each time we set different random positions $\bm{r}_\alpha$ and random energies $\epsilon_\alpha$.

We also look for the mean square displacement (MSD) of the exciton from the center of the mesa structure, where it was initially created,
\begin{equation}
    \bigl\langle r^{2}(t) \bigr\rangle = \biggl\langle \sum_\alpha r_\alpha^{2} \bigl|c_\alpha(t)\bigr|^2 \biggr\rangle,
    \label{eq:MSD}
\end{equation}
where $r_\alpha \equiv |\bm{r}_{\alpha0}| = |\bm{r}_\alpha - \bm{r}_0|$ is the distance from the QD initially occupied ($\alpha=0$).

In addition to that, we model the temporal dependence of the PL intensity.
In each of the $N_\mathrm{rep}$ realizations, we record the timestamp of the emission jump and, on the basis of that, we form a histogram of a number of jumps in each time period $\Delta t$.
The PL intensity is proportional to
\begin{equation}
    I \propto \frac{\langle N_\mathrm{ex} \rangle}{N_\mathrm{rep} \Delta t},
    \label{eq:photoluminescence}
\end{equation}
where $\langle N_{\mathrm{ex}} \rangle$ is the number of light quanta emitted in the time $[t,t+\Delta t)$ averaged over the realizations.
The time interval varies appropriately to achieve equal spacing on a logarithmic scale.

All the results obtained for the dissipative system employ the quantum jump method.
In contrast, for dissipationless systems, we use exact diagonalization to solve the unitary equation of motion given by the first term in~right-hand side of Eq.~\eqref{eq:master_equation}.

\section{Results of the numerical simulations \label{sec:numerical_results}}

\begin{figure}[t]
    \centering
    \includegraphics[width=\linewidth]{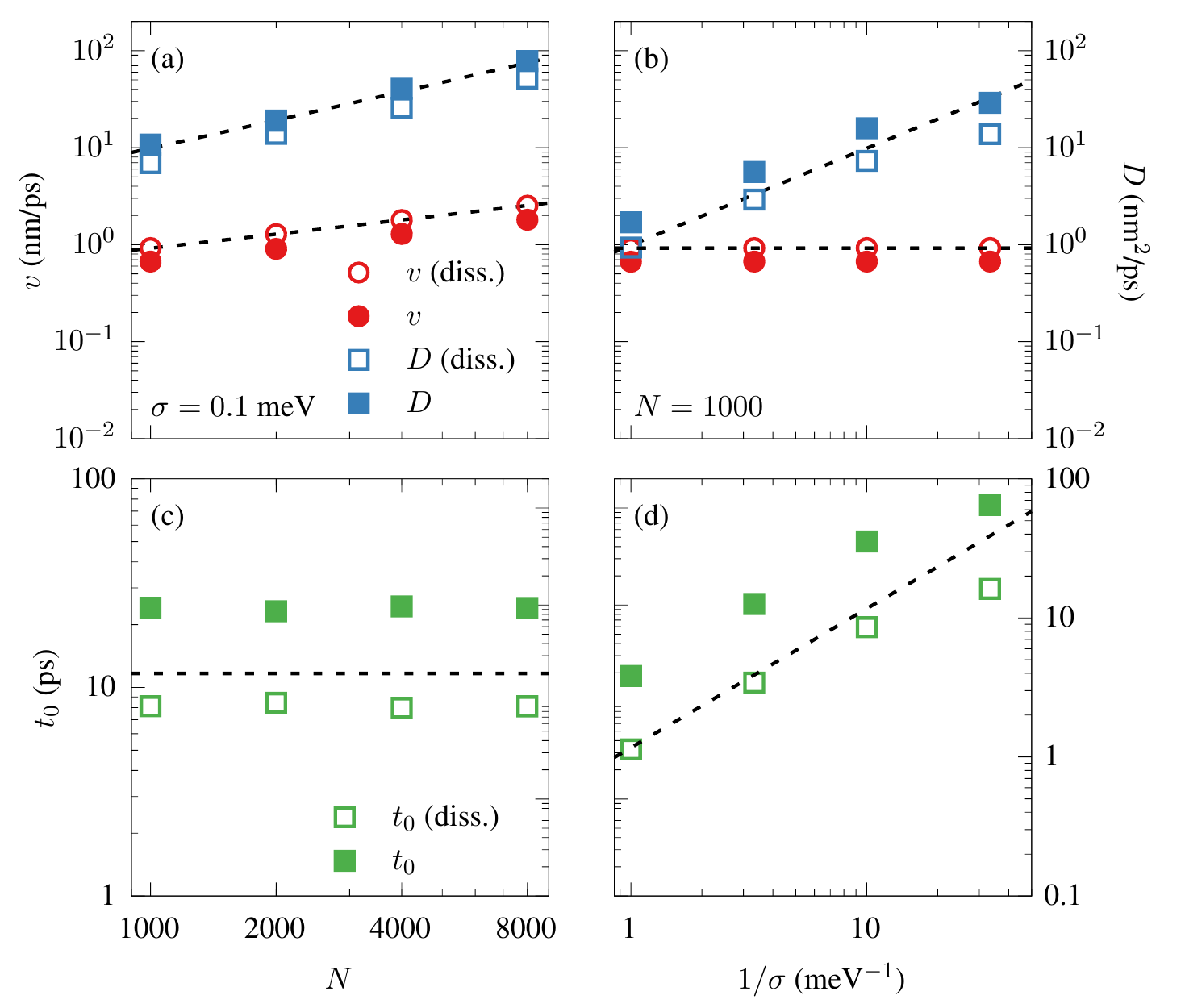}
    \caption{(a,b) Velocity of the ballistic motion (in red) and diffusion coefficient of normal diffusion (in blue) as a function of the system size and inverse disorder respectively; (c,d) crossover time between ballistic and diffusive stages (in green) as a function of the system size and inverse disorder strength, respectively.
    Open symbols correspond to the dissipative system whereas solid symbols to the idealized nondissipative system. Dashed lines respond for the approximate analytical results, see Sec.~\ref{sec:analytic}.}
    \label{fig:v_D_t0}
\end{figure}

In this section, we present the results obtained by a numerical implementation of the~model presented in~Sec.~\ref{sec:system_and_model}.
Specifically, we present the temporal evolution of MSD [Eq.~\eqref{eq:MSD}] comparing the system of a realistic limited exciton lifetime with an idealistic non-dissipative system.
We also present the spatial distribution of exciton occupations $\mathrm{X}_\mathrm{num} = \left\langle\abs{\braket{\Psi}{\Psi}}^2 \right\rangle$ in the ensemble and reveal the PL intensity.

\subsection*{Three-step dynamics}
Fig.~\ref{fig:evolution} shows MSD [Eq.~\eqref{eq:MSD}] as a function of time for dissipating systems [panels (a) and (b)] and dissipation-free systems [panels (c) and (d)] for several values of system size $N$ and standard deviation $\sigma$ of fundamental transition energy mismatch $\epsilon_\alpha$.
Straight lines on a doubly logarithmic scale indicate the power-law dependencies of the MSD on time.
MSD evolves in three subsequent steps.
First, for very short times, we observe ballistic transport with constant velocity~$v$,
\begin{equation}
    \bigl\langle r^{2}(t) \bigr\rangle_{r,\epsilon} = v^2t^2, \quad t<t_0.
    \label{eq:ballistic}
\end{equation}
Then, at a certain point in time $t_0$, standard diffusion with an appropriate diffusion constant $D$ starts,
\begin{equation}
    \bigl\langle r^{2}(t) \bigr\rangle_{r,\epsilon} = Dt, \quad t_0 < t < t_1.
    \label{eq:diffusive}
\end{equation}
Finally, at some time $t_1$, MSD saturates.
\begin{equation}
    \bigl\langle r^{2} (t) \bigr\rangle_{r,\epsilon} = r_\mathrm{sat}^2,\quad t_1<t,
    \label{eq:saturation}
\end{equation}
which is only visible in an idealized system with an infinite exciton lifetime [Fig.~\ref{fig:evolution}(c,d)].
Otherwise, the exciton decays before saturation is reached.
From the requirement for the MSD to be a continuous function, one gets the crossover time between the ballistic and diffusive stages,
\begin{gather}
    t_0 = D/v^2,
\end{gather}
as well as the crossover time between the standard diffusion and saturation stages,
\begin{gather}
    t_1 = r_\mathrm{sat}^2/D.
\end{gather}

\begin{figure}[t]
    \centering
    \includegraphics[width=\linewidth]{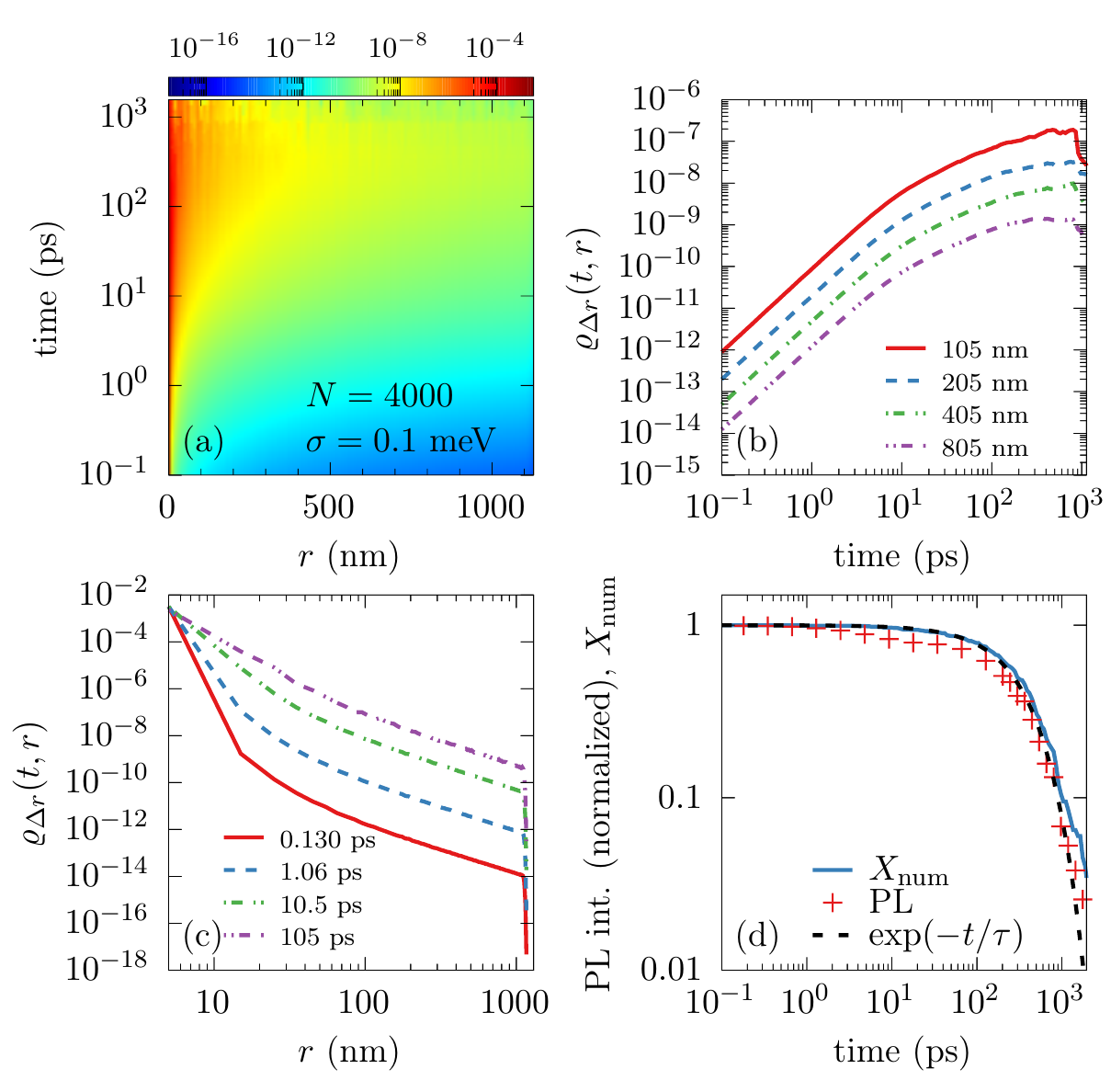}
    \caption{(a) Color-map of an occupation density $\varrho_{\Delta r}(t,r)$ as a function of time and distance from the central QD.
    For the sake of clarity, the data was truncated not to cover noise points after exciton recombination ($t\gtrsim 4\tau$);
    (b) Temporal evolution of $\varrho_{\Delta t}(t,r)$ for a few values of the distance from the center [the lines correspond to the vertical sections of the panel (a)].
    (c) Distance dependence of $\varrho_{\Delta t}(t,r)$ for a few values of time [the lines correspond to the horizontal sections of the panel (a)].
    (d) Photoluminescence intensity (see Eq.~\eqref{eq:photoluminescence}) and exciton decay in time.
    The dashed line represents the ideal exponential decay with decay time $\tau = 390$~ps as assumed in the simulations.}
    \label{fig:map}
\end{figure}

\subsection*{Dynamical parameters}

We found the dynamical parameters of diffusion, that is, the ballistic velocity $v$ and the diffusion coefficient $D$ (for dissipative and idealized systems) together with the diffusion range $r_\mathrm{sat}^2$ (only for idealized systems), and crossing times $t_0$ and $t_1$ by fitting the appropriate power functions to the corresponding stages of motion [Eqs.~\eqref{eq:ballistic}--\eqref{eq:saturation}].
In addition, the saturation level can be found directly from the exact diagonalization of \eqref{eq:hamiltonian_atomic_plus_coupling} which we explain in Appendix~\ref{appendix}.

In~Fig.~\ref{fig:v_D_t0} we present the dependence of~the~velocity and~diffusion coefficient on~the~size of~the~system and the strength of~the~energy disorder.
The dependencies seem to follow power laws with an~integer or~a~simple rational exponent.
The velocity grows as the square root of~the~size of~the~system [$v\propto N^{0.481\pm 0.011}$ from the fitting; see Fig.~\ref{fig:v_D_t0}(a)] and is~independent of~the~disorder [Fig.~\ref{fig:v_D_t0}(b)].
The diffusion coefficient increases linearly with the size of the system [$D\propto N^{0.969 \pm 0.037}$ from fitting, see~Fig.~\ref{fig:v_D_t0}(a)], but decreases with~the~disorder strength as $\propto 1/\sigma$ [$D \propto \sigma^{-0.955 \pm 0.012}$ from fitting, Fig.~\ref{fig:v_D_t0}(b)] at least for a strong disorder.
As the disorder decreases, the diffusive stage becomes less and less visible.
It is reflected in the dependence of diffusion coefficient $D$ in Fig.~\ref{fig:v_D_t0}(b), which deviates from the trend $1/\sigma$ for decreasing disorder (right part of the panel).

Dissipation changes the values of the ballistic velocity and the diffusion coefficient.
In Fig.~\ref{fig:v_D_t0}(a) and Fig.~\ref{fig:v_D_t0}(b) one notices that the ballistic velocity in the dissipative system is larger by a constant multiplicative factor compared to the velocity in the idealized system.
It follows that dissipation leads to an acceleration of diffusion in its first stage.
This means a faster emptying of the central QDs in favor of the other dots in the system.
More specifically, the enhancement of ballistic motion stems from the contribution of the dissipator matrix element $\Gamma_{\alpha\beta}$ to the coupling magnitude, which we show in Sec.~\ref{sec:analytic}.
On the contrary, in the second stage of motion, the diffusion coefficient in the idealized system exceeds the diffusion coefficient in the dissipative system by a constant multiplicative factor [Fig.~\ref{fig:v_D_t0}(a) and Fig.~\ref{fig:v_D_t0}(b)].
The physical reason for this can be explained as follows.
The form of the effective potential (Eq.\eqref{eq:effective-potential-H-function} and Fig.~\ref{fig:effective-potential-H-function}) causes the localized states to dissipate more slowly than the extended ones because of a negative imaginary part of $V^\mathrm{eff}_{\alpha\beta}$ in the region close to the initially excited QD.
Consequently, as evolution progresses, the proportion of the localized states compared to the extended ones increases, suppressing diffusion.

\subsection*{Exciton spacial distribution}

\begin{figure}[tb]
    \centering
    \includegraphics[width=\linewidth]{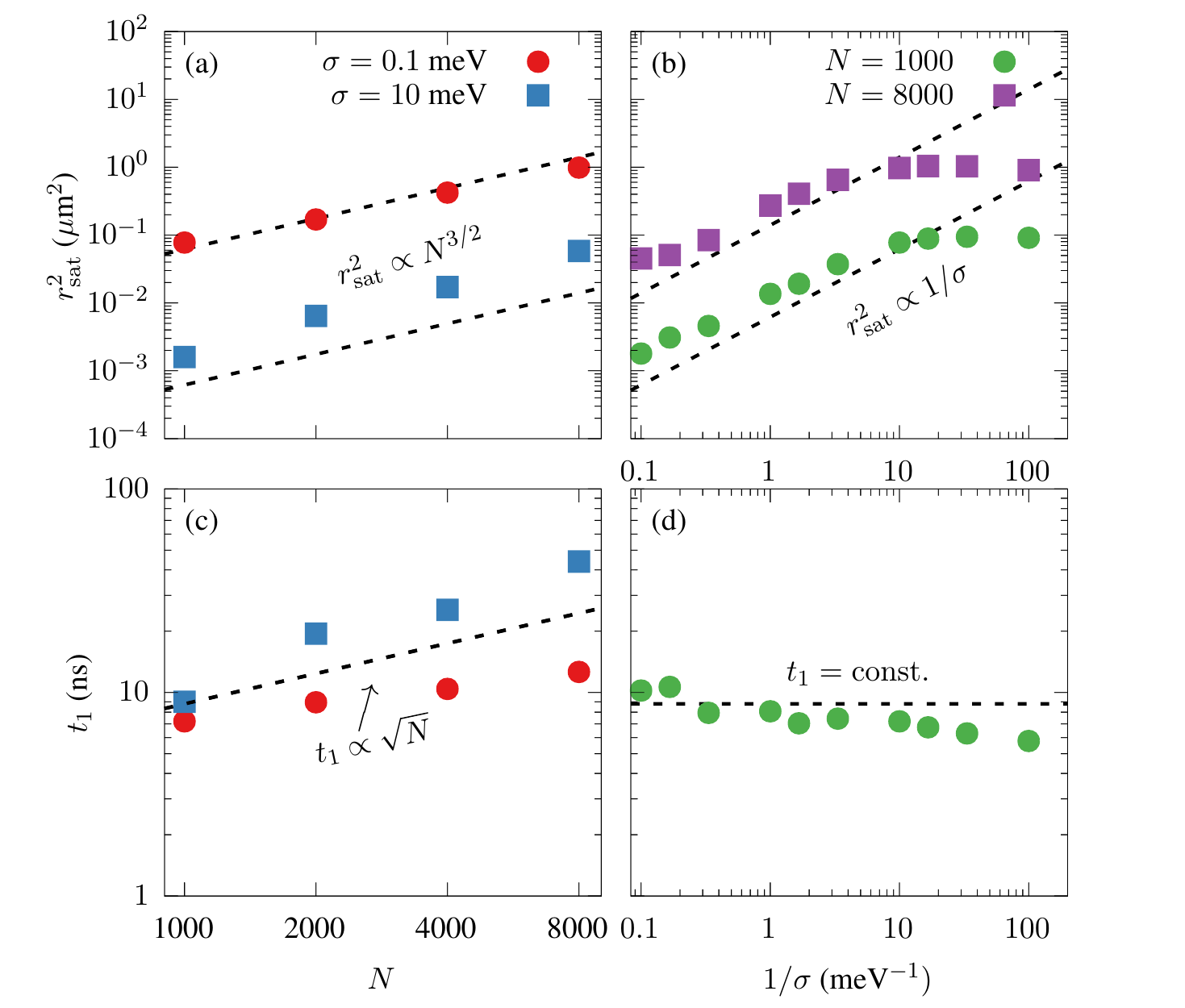}
    \caption{(a,b) The level of saturation of mean square displacement as a function of the size of the system (a) and inverse disorder strength (b), respectively. (c,d) Crossover time, at when the saturation stage begins, similarly as a function of the size of the system (c), and inverse disorder strength (d), respectively. The data was obtained mostly by numerical fit to MSD. Only the violet squares in panel (b) were obtained via exact diagonalization method explained in Appendix~\ref{appendixB}}
    \label{fig:saturation}
\end{figure}

In Fig.~\ref{fig:map}(a), we present a color map of the temporal and spatial dependence of the occupation density $\mathcal{\varrho}_{\Delta r}(t,r)$.
Separately, we illustrate how the occupation density at certain distances grows in time [Fig.~\ref{fig:map}(b)] and how occupations are distributed in space for several values of time [Fig.~\ref{fig:map}(c)].
The data are presented for a dissipative system.
The occupation at each length follows three steps similar to MSD: first, quadratic in time, then linear in time, and finally saturates.
The spatial decay of $\varrho_{\Delta r}(t,r)$ can be considered as a power law with different power exponents: higher for close QDs than for remote ones.
In Table~\ref{tab:fitting} we gather these exponents in the long-distance regime, corresponding to the subsequent curves in Fig.~\ref{fig:map}(c).
The value of the exponent is close to $-2$, which is consistent with the analytical solution presented in Sec.~\ref{sec:analytic}.

\begin{table}[b]
    \centering
    \begin{tabular}{c c}
    \hline\hline
    time (ps) & exponent $\mu$\\
    \hline
        $0.13$ &  $-2.1147 \pm 0.0098$\\
        $1.06$ & $-2.1213 \pm 0.0097$\\
        $10.54$ & $-2.1715 \pm 0.0089$ \\
        $104.82$ & $-2.194 \pm 0.022$ \\
        \hline\hline
    \end{tabular}
    \caption{The exponent of the power-law distribution of the exciton occupation density [see Fig.~\ref{fig:map}(c)] in the limit of long distance from the central QD, $\propto r^{\mu}$.}
    \label{tab:fitting}
\end{table}

\subsection*{Diffusion range}
It is clear from Fig.~\ref{fig:evolution} that the exciton life is too short to reach saturation.
Therefore, we determine the extent of diffusion for the idealized non-dissipative system.
The saturation level shown in Fig.~\ref{fig:saturation}(a) increases with the number of dots in the system as $\propto N^{3/2}$, which implies an increase as $\propto R^3$.
This trend cannot continue for large systems because the MSD increases faster than $\propto R^2$.
This may mean that the exciton will reach the mesa border for some value of $N$ and the subsequent growth will continue with the trend $\propto R^2$.
This change should be visible for homogeneous systems; however, even for $\sigma = 0.1$~meV the trend is still $N^{3/2}$ [Fig.~\ref{fig:saturation}(a), see last paragraph of Section~\ref{sec:analytic} for a discussion].

The diffusion range of a single exciton is about $100$~nm [Fig.~\ref{fig:evolution}(a,b)], which corresponds to a time of several nanoseconds.
Excitation spreads throughout the system, as shown in Fig.~\ref{fig:map}(a), the population of distant dots is small, less than $10^{-8}$.
However, in the $100$~nm region around the excitation center a few nanoseconds after excitation, the occupancies remain at the level of $\sim 10^{-4}$.
The PL intensity shown in~Fig.~\ref{fig:map}{(d)} as a~function of time is still not negligible for~this time-space regime.

\subsection*{Photoluminescence \& occupation decay}
The time dependence of the PL intensity and the total exciton occupation are presented in Fig.~\ref{fig:map}(d).
Both exhibit nearly exponential decay.
The numerical fit of the function $\exp(-t/\tau)$ in an interval $[0,2\tau]$ gives an exciton decay time of $\tau_X=(435.2\pm2.5)$~ps and PL decay time $\tau_{\text{PL}} = (279\pm12)$~ps.
Therefore, from the beginning of evolution, while exciton decays slower, PL decays faster than the independent emitter ($\tau=390$~ps).
However, at later times, the PL intensity slows down, even below the independent QD rate seen in Fig.~\ref{fig:map}(d) (red crosses on the right-hand side of the panel).
In Ref.~\cite{Kozub2012EnhancedCoupling++} it was suggested that the enhanced emission is caused by short-range couplings due to tunneling and Coulomb correlations.
Long-range couplings are too weak to impact emission in a strongly disordered ensemble.
However, in the case of a homogeneous system ($0.1$~meV in Fig.~\ref{fig:map}) the F\"orster couplings start to play a role, which is visible in the change of the ensemble emission against the emission of the independent emitters.

\subsection*{The central atom model}
In Fig.~\ref{fig:evolution}(c) dashed lines indicate the results of numerical simulations made in the first-order approximation (which we refer to as the central atom model), in which only direct couplings from central QD are relevant, while the other are set to zero.
The data follow the solution of the full model especially in the ballistic and diffusive stages of motion and deviate slightly in the saturation regime.
This approximation is justified when the strength of the disorder is much greater than the F\"orster coupling between QDs and the size of the system is not too big.
The central atom model allows one to analytically approximate the full model solution, which is the subject of the next section.

The central atom model works well for high disorder.
In fact, it is the first-order approximation of the Anderson series (locator expansion) as introduced in Ref.~\cite{Anderson1958}, which works in a high-disorder regime.
From a physical point of view, the energy disorder of $0.1$~meV corresponds to the uniform ensemble; however, compared to the magnitude of the F\"orster coupling, it is strong and, for that reason, operating within the central atom model is justified.

\section{Approximate analytical approach \label{sec:analytic}}
In this section, we present an approximate analytical solution to the model presented in Sec.~\ref{sec:system_and_model} that qualitatively reproduces the results of numerical studies obtained in Sec.~\ref{sec:numerical_results}.
This analytical approach was previously introduced in two different ways in~Refs.~\cite{Kawa2020DiffusionCoupling} and~\cite{Kawa2021SpreadOfCorrel} for a general lattice model of $N$ sites with long-range power-law coupling, $V(r)\propto 1/r$ in~Ref.~\cite{Kawa2020DiffusionCoupling} and, more generally, for $V(r)\propto 1/r^\mu$ in~Ref.~\cite{Kawa2021SpreadOfCorrel}.
Here, we extend that analytic approach to ensembles with randomly placed QDs and the oscillating three-term dipole coupling of Eqs.~\eqref{eq:coupling} and~\eqref{eq:coupling_function}.

\subsection*{Excitonic occupation}
Due to the symmetry of the central atom model, all QDs distant by $r$ from the center should have on average the same occupation $\langle | c_r(t) |^2\rangle$.
In Ref.~\cite{Kawa2020DiffusionCoupling} we have found it as an analytical solution of Anderson locator expansion \cite{Anderson1958} in the first order provided by the central atom model.
In Ref.~\cite{Kawa2021SpreadOfCorrel} instead, by exact diagonalization, we obtained an equal formula just considering direct transfer between two sites coupled via long-range coupling $V(r)$.
Here, we extend the latter strategy to a realistic dissipative system.
Let us consider Eq.~\eqref{eq:continuous-evolution-between-jumps} for two QDs separated by a distance of $r$ and with fundamental energy difference $\epsilon_r\in \mathcal{N}(0,2\sigma^2)$, which is the difference of two normally distributed random variables.
The set of equations of motion for the amplitudes $c_0(t)$ and $c_r(t)$ of the two-element basis state vector $\ket{\Psi} = c_0(t)\ket{0} + c_r(t)\ket{1}$ [Eq.~\eqref{eq:continuous-evolution-between-jumps}] satisfying Eq.~\eqref{eq:ketPsi} takes the form
\begin{align}
    i\hbar \dot{c}_0 & = \frac{\hbar}{2i}\Gamma_0 c_0 + V^\mathrm{eff}(r) c_r,\\
    i\hbar \dot{c}_r & = V^\mathrm{eff}(r) c_0 + \left( \epsilon_r + \frac{\hbar}{2i}\Gamma_0 \right) c_r,\label{eq:układ rownań dla dwóch kropek}
\end{align} 
with initial conditions $c_0(0) = 1$ and $c_r(0)=0$.
$V^\mathrm{eff}(r)$ corresponds to the effective coupling defined by Eqs.~\eqref{eq:effective-potential-H-function} and \eqref{eq:H(x)}.
The exact integration of this set of equations is available and has been widely examined, e.g. in the context of dressed states in quantum optics \cite{Gerry_Knight_2004}.
The interesting occupation of the acceptor-QD is
\begin{equation}
    \bigl|c_r(t)\bigr|^2 = e^{-\Gamma_0 t /2} |V^\mathrm{eff}(r)|^2 h(t),
    \label{eq:occupation}
\end{equation}
where
\begin{equation}
    h(t; \Omega) = \frac{\sin[{\Omega t/(4\hbar)}]}{(\Omega/4)^2},
\end{equation}
and $\Omega = \sqrt{\epsilon_r^2 + 4|V^\mathrm{eff}(r)|^2}$.
Since the quantum jump method demands normalization of the state vector until the jump takes place, the dissipative factor $e^{-\Gamma_0 t/2}$ must be neglected and set to one.

\begin{figure}[t]
    \centering
    \includegraphics[width=\linewidth]{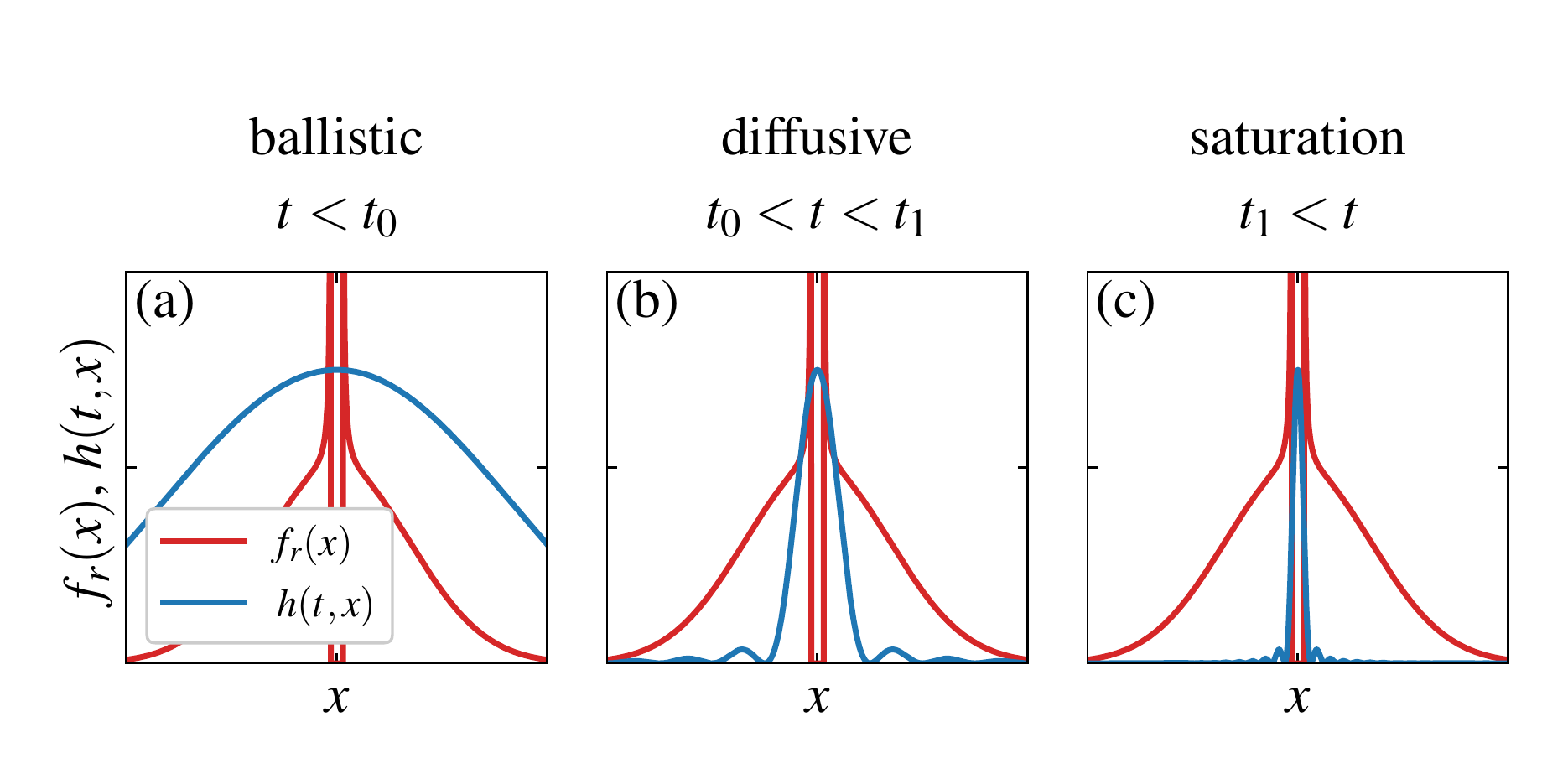}
    \caption{The probability density function (PDF) of eigenvale separation in the central atom model (the same red curve in each panel) versus the function $h(t,x)$ of Eq.~\eqref{eq:occupation} (blue curves).
    Panels correspods to ballistic (a), diffusive (b) and stauration (c) stages of evolution.}
    \label{fig:distribution}
\end{figure}

The average over the distribution of fundamental transition energies can be evaluated as an integral with the probability density function (PDF) $f_r(x)$ for the eigenvalue separation $\Omega$,
\begin{align}
    \bigl\langle |c_r(t) |^2 \bigr\rangle = |V^\mathrm{eff}(r)|^2 \int\limits_{-\infty}^\infty dx  f_r(x) h(t;x) \dd x.
    \label{eq:averaged-occupation}
\end{align}
The function $f_r(x)$ is close to the PDF of diagonal energy separation $f_\infty(x)$, which is a normal distribution of zero mean and standard deviation $\sqrt{2}\sigma$, but includes a narrow gap around zero of width $4|V^\mathrm{eff}(r)|$ reflecting levels repulsion (see Fig.~\ref{fig:distribution}) \cite{Kawa2020DiffusionCoupling,Kawa2021SpreadOfCorrel}.
The index ``$\infty$'' refers to the infinite distance between QDs that corresponds to the lack of coupling, which also implies $\Omega = \epsilon_r$.
In the presence of coupling, two sites of bare energy separation $\epsilon_r$ contribute to two eigenvalues separated by $\Omega = \sqrt{\epsilon_r^2 + 4|V^\mathrm{eff}|^2}$.
Taking advantage of this, we find that
\begin{align}
\begin{aligned}    
    f_r(x) &= \frac{\dd}{\dd x} \int\limits_{-\infty}^{\sqrt{x^2 - 4|V^\mathrm{eff}|^2}} \dd \overline{x} f_\infty(\overline{x})\\
    &= f_{\infty}\left(\sqrt{x-4|V^\mathrm{eff}(r)|^2}\right) \frac{\abs{x}}{\sqrt{x^2-4|V^\mathrm{eff}(r)|^2}}
\end{aligned}
\end{align}
for $\abs{x}>2|V^\mathrm{eff}(r)|$ and zero otherwise.

\subsection*{Mean square displacement}
For systems with spherical symmetry, MSD [Eq.~\eqref{eq:MSD}] can be expressed as
\begin{equation}
    \bigl\langle r^2(t) \bigr\rangle = \biggl\langle \sum_r r^2  \sum_{k=1}^{n_r} \bigl|c_r^{(k)}(t)\bigr|^2 \biggr\rangle,
    \label{eq:MSD_suma_po_pierscieniach}
\end{equation}
where the first sum runs through distances from the center, and the second sum passes over the $n_r$ QDs lying on a thin ring of radius $r$ in the ensemble.
In the continuous limit of spatial distribution QDs with constant surface density $\rho_A$, Eq.~\eqref{eq:MSD_suma_po_pierscieniach} takes the form of
\begin{align}
\bigl\langle r^2(t) \bigr\rangle=2\pi \rho_A \int\limits_{0}^{R}  r^3 \bigl\langle | c_r(t) |^2 \bigr\rangle \dd r,
\label{eq:MSD_integral_formula}
\end{align}
where the integration covers the circular area of the mesa of radius $R$.
The index $k$ can be omitted due to the symmetry of the central atom model, that is, all QDs distant from the center by $r$ should have (on average) the same occupation.
Evaluation of \eqref{eq:averaged-occupation} and then \eqref{eq:MSD_integral_formula} is based on dividing the evolution into three time regimes: for very short times, for moderate times, and finally for long times.

\subsection*{Ballistic motion}
For very short times, $t < \hbar/\sigma$, $h(t;x)$ is broad and slowly varying [cf. Fig.~\ref{fig:distribution} (a)], so it can be expanded into a Taylor series to the first order around $t=0$, which yields $h(x)\approx t^2/\hbar^2$.
In this case, the internal gap in the distribution does not play an important role and can be neglected, thus $f_r(x) \approx f_\infty(x) = \exp[-x^2/(4\sigma^2)]/\left(2\sqrt{\pi}\sigma\right)$ and the integration in \eqref{eq:averaged-occupation} yields
\begin{equation}
    \bigl\langle|c_r(t)|^2\bigr\rangle = |V^\mathrm{eff}(r)|^2 \frac{t^2}{\hbar^2}  \quad \text{for} \quad t<t_0.
    \label{eq:occupation_first_stage}
\end{equation}
Furthermore, we evaluate \eqref{eq:MSD_integral_formula}, which takes the form
\begin{align}
    \bigl\langle r^2(t) \bigr\rangle = \left(\frac{2\pi\rho_A}{\hbar^2} \int r^3 \langle |V^\mathrm{eff}(r)|^2 \rangle \dd r \right)t^2,
\end{align}
where the expression in parentheses is the squared ballistic velocity [Eq.~\eqref{eq:ballistic}],

\begin{align}
\begin{aligned}
    v^2 =& \frac{9}{64}\frac{\Gamma_0^2}{k_0^2} N + \frac{27}{32}\frac{\Gamma_0^2 \pi \rho_A}{k_0^4}\ln(\frac{R}{r_\mathrm{av}})\\
    &+ \frac{9}{64}\frac{\Gamma_0^2 \pi \rho_A}{k_0^4}\left(\frac{1}{k_0^2 r_\mathrm{av}^2} - \frac{1}{k_0^2R^2}\right).
\end{aligned}
\label{eq:v2-analitycznie}
\end{align}

\subsection*{Standard diffusion}
As time grows, $h(t;x)$ narrows.
At moderate times, $\hbar / \sigma < t < \hbar / \sqrt{|V^\mathrm{eff}(R)|^2}$, $h(x)$ becomes proportional to the unormalized Dirac delta of the area $2\pi t/\hbar$ but is still wide enough to be insensitive to the gap in $\langle f_r(x) \rangle_r$.
Thus, we again approximate $\langle f_r(x)\rangle_r \approx f_\infty(x)$ and obtain the occupation,
\begin{align}
\begin{aligned}
\bigl\langle|c_r(t)|^2\bigr\rangle &= |V^\mathrm{eff}(r)|^2 \int\limits_{-\infty}^\infty \dd x f_\infty(x) 2\pi \delta(x)/\hbar \\
&=|V^\mathrm{eff}(r)|^2\frac{\sqrt{\pi} t}{\hbar\sigma} \quad \text{for} \quad t_0 < t < t_1.
\end{aligned}
\label{eq:occupation_second_stage}
\end{align}
Again we evaluate Eq.~\eqref{eq:MSD_integral_formula} and obtain MSD for the diffusive stage of motion [Eq.~\eqref{eq:diffusive}], with the diffusion coefficient,
\begin{align}
    D = \frac{\sqrt{\pi}\hbar}{\sigma}v^2 \propto \frac{N}{\sigma}.
    \label{eq:D-analitycznie}
\end{align}
The crossover time between the first and the second stage is
\begin{align}
    t_0 = \frac{\sqrt{\pi}\hbar}{\sigma}.
\end{align}
Thus, it depends only on the magnitude of the fundamental transition energy disorder and is independent on the size or spatial density of the QD ensemble.

Analytical formulas for $v$, $D$, and $t_0$ are represented by a dashed line in Fig.~\ref{fig:v_D_t0}.
They provide trends that are at least qualitatively aligned with the numerical data.

\subsection*{Saturation. The diffusion range}
Finally, we find the diffusion range expressed by the saturation level of MSD.
For distant times, $t>\hbar/|V^\mathrm{eff}(R)|$, the central peak of $h(x)$ falls into the gap inside $f_r(x)$.
The peak is canceled, and the remaining part of $h(x)$ can be approximated by $h(x) \approx 1/(2x^2)$ where the oscillating nominator was averaged over its period.
We evaluate \eqref{eq:MSD_integral_formula} and obtain
\begin{equation}
    \begin{split}
        \bigl\langle | &c_r(\infty) |^2 \bigr\rangle= \\
        &=\frac{\sqrt{\pi}}{2\sigma}|V^\mathrm{eff}(r)| \erfc\left(\frac{|V^\mathrm{eff}(r)|}{\sigma}\right)\exp\left(\frac{|V^\mathrm{eff}(r)|^2}{\sigma^2}\right)\\
        &\approx \frac{\sqrt{\pi}}{2\sigma}|V^\mathrm{eff}(r)|\quad\text{for}\quad t_1<t.
    \end{split}
    \label{eq:occupation_third_stage}
\end{equation}
Next, we calculate the diffusion range using Eq.~\eqref{eq:MSD_integral_formula} which takes the form
\begin{equation}
    r^2_\mathrm{sat} = \frac{\pi^{3/2} \rho_A}{\sigma} \int r^3 |V^\mathrm{eff}(r)|\dd r.
    \label{eq:MSD-integral-saturation}
\end{equation}
The integrand in Eq.~\eqref{eq:MSD-integral-saturation} is a square root of a polynomial of $r$.
In general, the integral can be expressed using elliptic functions of the first and second types.
However, such a result is impractical and it is difficult to extract a trend in $R$ from it.
For simplicity, let us approximate the integral in Eq.~\eqref{eq:MSD-integral-saturation} by keeping only the largest term in the integrand, that is, proportional to $r^2$ (in the regime of large ensembles, $R\gg 1/k_0$).
Then we obtain
\begin{align}
\begin{aligned}
    r^2_\mathrm{sat} =& \frac{\pi^{3/2}\rho_A}{8}\frac{\hbar\Gamma_0}{\sigma k_0} R^3 =  \frac{\hbar\Gamma_0}{8\sqrt{\rho_A}k_0} \frac{N^{3/2}}{\sigma}\\
    &\approx 0.195 \frac{N^{3/2}}{\sigma [\text{meV}]} [\text{nm}^2].
\end{aligned}
    \label{eq:saturation_analytic_trend}
\end{align}
This result provides a growth of the diffusion range as $N^{3/2}$, which is consistent with the simulation results in Fig.~\ref{fig:saturation}(a,b).
The diffusion range grows faster than $R^2 \propto N$, which means that the exciton should reach the mesa border at some large $N$.
However, according to Eq.~\eqref{eq:saturation_analytic_trend} this may happen for systems of more than $5\cdot10^4$ QDs, which is far beyond the simulation possibilities.
In the realistic model, the saturation phase is most often not present in evolution because the exciton has already dissipated from the system [Fig.~\ref{fig:evolution}(a,b)].
Thus, we can only compare the analytical formula \eqref{eq:saturation_analytic_trend} with the results of the simulation of the idealistic model.

\section{Conclusions and discussion\label{sec:discussion}}

We have investigated the diffusion of an exciton in a planar, energetically inhomogeneous ensemble of randomly distributed QDs coupled by dipole interactions. We have shown that diffusion takes place in three stages: ballistic diffusive and saturation.
In each of these stages, the occupations are distributed according to power laws as a function of the distance from the initially excited QD.
Qualitatively, the dynamics is the same as in the generic lattice model studied in Ref.~\cite{Kawa2020DiffusionCoupling}.
This means that neither the random spatial distribution of the QDs nor the full structure of the dipole coupling, including the spatially oscillating factors, leads to essential corrections compared to the simple power-law coupling. 

The power-law coupling model is formally restored in the limit of large distances, when the leading term in the dipole coupling dominates, and for dense ensembles, when the oscillations in the coupling average out.
If, additionally, the energetic disorder is strong compared to the couplings, one can replace the full model by the ``central atom'' model, in which only the initially excited QD is coupled to all other QDs in the system.
In this case, we were able to derive an analytical solution that correctly reproduces the parametric dependences of the ballistic speed and diffusion coefficient on the system size and QD density. Quantitatively, the predictions of this model slightly differ from the simulations, which may be due either to the limitations of the ``central atom'' model (contribution from higher-order transitions) or to the approximations made to the coupling. 

Although our discussion referred to QD ensembles, the conclusions are valid for any system in which excitation can be transferred via dipole couplings, as long as they belong to the same parametric class of large system sizes (compared to resonant wavelength) and strong disorder (compared to dipole couplings at typical distances between QDs).

In Ref.~\cite{DeSales2004} the authors considered two different surface densities of QDs, which provided an important insight into the diffusion mechanism.
In our work, we considered a system with only one fixed surface density, for which the average distance between dots ($\approx 32$~nm) is large enough to allow us to neglect short-range couplings (e.g. tunneling) between QDs.
Any smaller surface density would be valid, and the observed difference in the effects would be purely quantitative.
This can be seen in the analytical formula for the total diffusion range [Eq.~\eqref{eq:saturation_analytic_trend}], where we present it also as a function of the surface density of the ensemble.
The diffusion range is inversely proportional to the square root of the QD surface density if the number of QDs in the ensemble is assumed constant.
This is consistent with the results presented in Ref.~\cite{DeSales2004}.
On the other hand, increasing the surface density would lead us beyond the utility of the employed model.
That is, if the QDs are close enough, the short-range effects start playing a role, and we also move towards the limit of validity of the dipole approximation.

FRET in QD ensembles remains a challenge for experimental analysis.
In the usual ensembles, where $\sigma$ is of the order of tens of meV, ballistic transport would shift to the femtosecond scale, making it hardly possible for experimental observation.
Regardless of the experimental recognition of the diffusion type, F\"orster transfer in QD ensembles remains a subtle effect, which may explain its limited documentation in the literature.

\section*{Acknowledgements}
This research is part of project no. 2021/43/P/ST3/03293 co-funded by the National Science Centre and the European Union’s Horizon 2020 research and innovation program under Marie Sklodowska-Curie grant agreement no. 945339.

Calculations were partially carried out using resources provided by Wroc\l{}aw Centre for Networking and Supercomputing~\footnote{\url{http://wcss.pl}}, grant No. 203.

\appendix

\begin{figure}[tb]
    \centering
    \includegraphics[width=\linewidth]{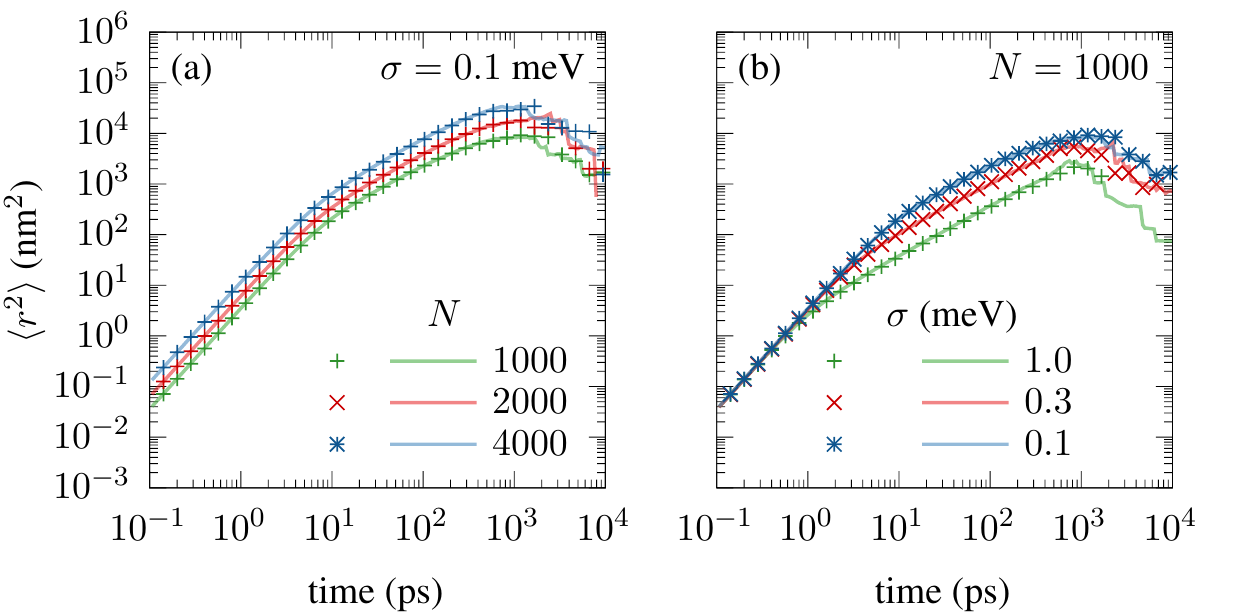}
    \caption{Mean square displacement of the exciton over time in the presence of spin-flipping F\"orster transfer, as detailed in Appendix~\ref{appendixB}.
    Solid lines denote scenarios without exciton FSS, while points represent results with non-zero FFS, specifically normally distributed with a standard deviation of $0.1$ meV in each case.
    Panel (a) represents data for varying system sizes with a fixed disorder of $\sigma=0.1$ meV.
    In panel (b), data are depicted for different strengths of the disorder with a constant number of quantum dots, $N=1000$. The number of disorder realizations for each dataset is consistent at $N_\mathrm{rep}=150$.}
    \label{fig:appendix-exchange}
\end{figure}

\section{Saturation level from direct diagonalization} \label{appendix}
The saturation level of the MSD can be calculated by fitting a constant function numerically to an MSD at late times.
Here, we show that the saturation level of the population, and thus for MSD, can be extracted directly from exact diagonalization without any fitting.
The initial state of the system corresponds to a fully occupied central QD.
Within this appendix, let us index that QD by $\alpha_0$.
The state of the system after some time is given by an action of evolution operator on the ket state $\ket{\alpha_0}$,
\begin{align}
    \ket*{\Psi(t)} = & e^{-i\hat{H}t}\ket*{\alpha_0} \nonumber \\ 
    = &\sum\limits_\alpha \left(\sum_n e^{-iE_nt} \braket*{\alpha}{n}\braket*{n}{\alpha_0}\right) \ket*{\alpha},\label{eq:wyprowadznie_ewolucja}
\end{align}
where $\ket{n}$ is an eigenket of $\hat{H}$ with energy $E_n$.
The expression in parentheses corresponds to the amplitude $c_\alpha(t)$ of Eq.~\eqref{eq:ketPsi}.
The corresponding occupation is given by
\begin{align}
    \abs{c_\alpha(t)}^2 = \sum\limits_{n,m} e^{-i(E_n-E_m)t} \braket*{\alpha}{n}\braket*{n}{\alpha_0}\braket*{m}{\alpha}\braket*{\alpha_0}{m}.
\end{align}
In the limit of infinite time, only the term with $E_m=E_n$ is important, and the saturation level of occupation takes the form of
\begin{equation}
    \abs{c_\alpha(\infty)}^2 = \sum\limits_n \abs{\braket*{\alpha}{n}}^2 \abs{\braket*{\alpha_0}{n}}^2.
\end{equation}

\section{Exciton fine structure splitting and spin-flipping F\"orster transfer} \label{appendixB}
The model used for the description of diffusion in this work involves some simplifications.
In particular, exciton polarization was assumed to be preserved, and thus spin-flipping F\"orster transfer was neglected.
However, QDs support two bright exciton states.
First, with total angular momentum $j_z=1$ represented as $\ket{\mathrm{e},\mathrm{hh}} = \ket{\downarrow, \Uparrow}$ and the other with $j_z=-1$, represented as $\ket{\mathrm{e},\mathrm{hh}} = \ket{\uparrow, \Downarrow}$, where ``$\mathrm{e}$'' stands for electron, and ``$\mathrm{hh}$'' stands for heavy hole state.
Thus, both excitons can be addressed by circularly polarized light.
As shown in Ref.~\cite{Specht2015} the F\"orster transfer enables spin-preserving and spin-flipping transfers with different magnitudes.

In addition, on-site exciton fine-structure splitting (FSS) can disturb the diffusion process, leading to Rabi oscillations between the bright exciton states within each QD.
In this appendix, we present an extended model, taking into account the exciton FSS and two types of F\"orster transport --- spin preserving and spin flipping.
The derivation of such a model is straightforward extension, where now we consider dipole operators for each bright exciton separately, that is, $\bm{\hat{d}}_\alpha^+ = \bm{d}_0^+(\dyad{+}{0})_\alpha + \mathrm{H.c.}$ and $\bm{\hat{d}}_\alpha^- = \bm{d}_0^-(\dyad{-}{0})_\alpha + \mathrm{H.c.}$
As substituted for Eq.~\eqref{eq:coupling_PZW_transform} and following the steps of the derivation of the equation of motion presented in Ref.~\cite{Lehmberg1970RadiationSystem}, one obtains the spin-preserving F\"orster couplings $V_{\alpha\beta}^{aa}=-\hbar\Gamma_0G(k_0r_{\alpha\beta})$ identical to Eq.~\eqref{eq:coupling} and spin-flipping ones $V_{\alpha\beta}^{a\bar{a}}=-\hbar\Gamma_0 G^{a\bar{a}}(k_0r_{\alpha\beta})$ for $\alpha\neq\beta$ with
\begin{equation}
    G^{a\bar{a}}(x) = \frac{3}{8} e^{2i\varphi a}\left[ \frac{\cos x}{x} - 3\left( \frac{\sin x}{x^2} + \frac{\cos x}{x^3} \right) \right],
\end{equation}
where the geometrical phase $\varphi$ is attained.
In addition, the diagonal term (in lattice indices) $V_{\alpha\alpha}^{a\bar{a}}$ corresponds to the exciton FFS, which we model here phenomenologically as a normally distributed random variable of standard deviation $0.1$~meV.
The index $a=\pm$ corresponds to the exciton polarization and $\bar{a}$ denotes the polarization opposite to $a$.
Similarly, the elements of the dissipator matrix for the spin-preserving coupling are $\Gamma_{\alpha\beta}^{aa} = \Gamma_0F^{aa}(k_0r_{\alpha\beta})$, where $F^{aa}(x)\equiv F(x)$ defined in Eq.~\eqref{eq:dissipator-distance-dependence}, while for spin flip transitions $\Gamma_{\alpha\beta}^{a\bar{a}} = \Gamma_0 F^{a\bar{a}}\left(k_0 r_{\alpha\beta}\right)$ with
\begin{equation}
    F^{a\bar{a}}(x) = -\frac{3}{4}e^{2i\varphi a} \left[ \frac{\sin x}{x} + 3\left( \frac{\cos x}{x^2} - \frac{\sin x}{x^3} \right) \right].
\end{equation}
Since the transfer depends on the absolute value of the F\"orster coupling, it should not depend on the geometrical phase~$\varphi$.
The mean square displacement of the exciton moving within the model presented is shown in Fig.~\ref{fig:appendix-exchange} for some system sizes (a) and disorder strengths (b).
Initially, the cetral QD is excited by circularly polarized light into one of the two bright exciton states, say $\ket{\downarrow,\Uparrow}$.
The diffusion follows a three-step difusion as before.
Again, the saturation stage is poorly seen because of the finite exciton lifetime.
One can see points representing the data for non-zero FFS follow lines, which correspond to the lack of the FSS.
This suggests that the FSS barely affects the diffusion.

\end{document}